\documentclass[11pt]{article}
\usepackage{graphicx}
\begin{document}
%

\begin{center}
{\large \bf A precise definition of the Standard Model}

\vskip.5cm

W-Y. Pauchy Hwang\footnote{Correspondence Author;
 Email: wyhwang@phys.ntu.edu.tw}
 \\
{\em Asia Pacific Organization for Cosmology and Particle Astrophysics, \\
Institute of Astrophysics, Center for Theoretical Sciences,\\
and Department of Physics, National Taiwan University,
     Taipei 106, Taiwan}
\vskip.2cm


{\small(22 September 2014; Revised: 24 June 2016)}
\end{center}

\begin{abstract}
We declare that we are living in the quantum 4-dimensional Minkowski
space-time with the force-fields gauge-group structure
$SU_c(3) \times SU_L(2) \times U(1) \times SU_f(3)$ built-in from
the very beginning. From this overall background, we see the lepton
world, which has the symmetry characterized by $SU_L(2) \times U(1)
\times SU_f(3)$ - the lepton world is also called "the atomic
world". From the overall background, we also see the quark world,
which experiences the well-known $(123)$ symmetry,
i.e., $SU_c(3) \times SU_L((2) \times U(1)$. The quark world
is also called "the nuclear world".

The $3^\circ\,K$ cosmic microwave background (CMB) in our
Universe provides the evidence of that "the force-fields
gauge-group structure was built-in from the very beginning".
The CMB is almost uniform, to the level of one part in $10^5$,
reflecting the massless of the photons.

The lepton world is dimensionless in the 4-dimensional
Minkowski space-time. That is, all couplings are
dimensionless. The quark world is also dimensionless
in the 4-dimensional Minkowski space-time. Apart from
the "ignition" term, the gauge and Higgs sector, i.e.,
the overall background, is also dimensionless. Thus,
apart the "ignition" term, our world as a whole is
dimensionless in the 4-dimensional Minkowski space-time
- that is, it is the characteristic of the quantum
4-dimensional Minkowski space-time.

\bigskip

{\parindent=0pt PACS Indices: 12.60.-i (Models beyond the standard
model); 98.80.Bp (Origin and formation of the Universe); 12.10.-g
(Unified field theories and models).}
\end{abstract}

\bigskip

\section{Prelude}

What is the Standard Model? It is a model that describes the
behaviors of the point-like particles such as the electrons,
the photons, the quarks, etc. What is the Standard Model
of All Centuries \cite{Hwang417}? We believe that the
description of the point-like particles, the smallest units
of matter in our Universe, on the basis of the Einstein's
relativity principle and the quantum principle is so
fundamental and also so complete and so consistent, that
this Standard Model would stay there longer than the
Newton's classic era.

The entries in the quantum 4-dimensional Minkowski
space-time with force-fields gauge-group structure
$SU_c(3) \times SU_L(2) \times U(1) \times SU_f(3)$
built-in from the very beginning mean the objects
that have definitive properties both under the
force group $SU_c(3) \times SU_L(2) \times U(1)
\times SU_f(3)$ and under the Lorentz group
defining the quantum 4-dimensional Minkowski
space-time.

What precisely is the difference between "4-dimensional
Minkowski space-time with the force-fields gauge-group
structure $SU_c(3) \times SU_L(2) \times U(1) \times
SU_f(3)$ built-in from the very beginning" and just
"4-dimensional Minkowski space-time"? In the former,
every object must have the designated-group assignments
while, in the latter, it could be any 4-dimensional object
in the Minkowski space-time. Is the adjective necessary
in this context? We propose "yes" in the strictly
mathematical sense.

We have to do all these with extra care, since "the Standard
Model" is the beginning of everything. There exist basically
no spoken rules in writing the Standard Model. In the Newton's
classic era, different languages were developed for expressing
ideas and things in a consistent and complete manner. In my
search of the Standard Model, in the beginning it is not clear
what I was searching for and after a period of successful
searches I kept trying to spell out what the Standard Model
really is. What is in the abstract of this paper, from
the early version into the current one (completely revised),
is a good illustrative example. It is for the Standard Model
of All Centuries.

Back in the 20th Century, we didn't realized that there is
something special for the complex scalar fields in the quantum
4-dimensional Minkowski space-time. A complex
scalar field $\phi(x)$ is born to be self-repulsive, due to
the $\lambda (\phi^\dagger \phi)^2$ interaction with
the positive dimensionless coupling $\lambda$. With a
negative $\lambda$, the system will collapse. So, if
alone, {\it it cannot exist.}

For the two complex scalar fields, the attractive mutual
interaction $-2\lambda (\phi_1^\dagger \phi_2) (\phi_2^\dagger \phi_1)$
would be enough to overcome the self-repulsiveness of the two
individual complex scalar fields. $\lambda (\phi^\dagger \phi)^2$
in fact writes the story for everything.

The story which we put forward above about the {\it nonexistence} of
a single complex scalar field quite striking - whether it is right
or not is still waiting for a clear-cut mathematical proof. Basically,
$\lambda$ cannot be negative since the system would collapse to
the negative infinity. It cannot be zero since this would be
meta-stable. We suspect $\lambda={1\over 8}$ in our notations, but
we should admit that it is still lack of rigorous proof.

Another important question is this: Why do we have the gauge bosons
corresponding to the group $SU_c(3) \times SU_L(2) \times U(1)
\times SU_f(3)$ and, among these, why is only the photon massless?
We think that we are not ready to answer a question like this -
but eventually we would get enough hints for the final answer.

Among the force fields or gauge fields, most of the gauge fields,
upon spontaneous symmetry breaking (SSB), become massive, including
weak bosons and family gauge bosons, if the standard wisdom is
assumed. That calls for the Standard-Model (SM) Higgs $\Phi(1,2)$,
the purely family Higgs $\Phi(3,1)$, and the mixed family Higgs
$\Phi(3,2)$ - they interact attractively whenever possible, i.e.,
between $\Phi(1,2)$ and $\Phi(3,2)$, and between $\Phi(3,1)$
and $\Phi(3,2)$. (The two numbers/labels are referred to
$SU_f(3)$ and $SU_L(2)$ - that is, triplets, doublets, or
singlets.) Thus, these things provide the "background"
of everything else. This is what we mean by the
quantum 4-dimensional Minkowski space-time with the force-fields
gauge-group structure $SU_c(3) \times SU_L(2) \times U(1) \times
SU_f(3)$ imprinted at the very beginning.

So, if there would be only the SM Higgs $\Phi(1,2)$,
then {\it it cannot exist.} The self-repulsive
interaction $\lambda (\phi^\dagger \phi)^2$ (with
$\lambda > 0$) would make it from existence. As said
earlier, $\lambda=0$ would be meta-stable while
$\lambda<0$ would collapse. The question is why
it is $\lambda={1\over 8}$, and that should be
determined {\it globally} by the quantum 4-dimensional
Minkowski space-time, as this question doesn't arise
if {\it not} the quantum 4-dimensional Minkowski
space-time.

The algebra among the three Higgs $\Phi(1,2)$,
$\Phi(3,2)$, and $\Phi(3,1)$ arises {\it only}
when it is in the quantum 4-dimensional Minkowski
space-time. If the space-time differs from the
4-dimensional, the algebra simply doesn't apply.
It is rather strange!!

The next question associated with the Higgs fields is to understand
"the origin of mass" - a question that we have recently
gained some understanding \cite{Origin}. In that \cite{Origin}, we
may set all the mass terms of the various Higgs to identically zero,
except one spontaneous-symmetry-breaking (SSB) igniting term. All
the mass terms are the results of this SSB, when switched on.
Therefore, the "mass" is the result of SSB - a generalized Higgs
mechanism. Thus, when the temperature is higher than a certain
critical temperature, the notion of "mass" does not exist.

The set of the "various" Higgs includes the Standard-Model (SM) Higgs
$\Phi(1,2)$, the mixed family Higgs $\Phi(3,2)$, and the pure family
Higgs $\Phi(3,1)$, where the first label refers to the group $SU_f(3)$
while the second the group $SU_L(2)$. The ignition could be on the
pure family Higgs $\Phi(3,1)$ \cite{Origin}, and it is clear that
that it may not be on the SM Higgs $\Phi(1,2)$.

These related Higgs, being the scalar fields, act as the systems of
energies, self-interacting via dimensionless $\lambda
(\phi^\dagger \phi)^2$ and interacting equivalently with other
Higgs. When the temperature is low enough, it becomes the "mass"
phase, or the phase in which the particles have masses.

\medskip

\section{The Lepton World and the Quark World}

The overall background in our world is the quantum 4-dimensional
Minkowski space-time with the force-field gauge-group structure
$SU_c(3) \times SU_L(2) \times U(1) \times SU_f(3)$ imprinted
at the very beginning. It sees the lepton world, of atomic sizes.
It also sees the quark world, of much smaller nuclear sizes.

Thus, it is of importance to see that the
$SU_L(2) \times U(1) \times SU_f(3)$ symmetry is realized
in the lepton world, through the proposal
\cite{HwangYan} $((\nu_\tau,\,\tau)_L,\,(\nu_\mu,\,\mu)_L,
\,(\nu_e,\,e)_L)$ $(columns)$ ($\equiv \Psi(3,2)$) as the
$SU_f(3)$ triplet and $SU_L(2)$ doublet. It is also essential to complete
the Standard Model \cite{Hwang417} by working out the Higgs
dynamics in detail \cite{Origin}. Here it is important to realize the role
of neutrino oscillations - it is the change
of a neutrino in one generation (flavor) into that in another generation;
or, we need to have the coupling $i \eta \bar \Psi_L(3,2)\times
\Psi_R(3,1) \cdot \Phi(3,2)$, exactly the coupling introduced by Hwang
and Yan \cite{HwangYan}. Then, it is clear that the mixed family Higgs
$\Phi(3,2)$ must be there. The remaining purely family Higgs $\Phi(3,1)$
helps to complete the picture, so that the eight gauge bosons are
massive in the $SU_f(3)$ family gauge theory \cite{Family}. [For
consistency in our notations, it should be ${\tilde \Phi}(3,2)$ in the
paper of Hwang and Yan \cite{HwangYan}.]

With a complete Standard Model such as \cite{Hwang417}, we could address
a few basic questions. After all, all "building blocks of matter" seem to
be point-like particles (point-like Dirac particles if fermions), and vice
versa. And nothing more. If quantum field theory (QFT) can describe the
Nature, it should mean more - such as the various ultraviolet divergences,
do they cancel out in some way? See below on tests for a "complete" theory.

Usually in an old textbook \cite{Books}, the QCD chapter precedes the one
on Glashow-Weinberg-Salam (GWS) electroweak theory, but we are talking
about the $SU_c(3) \times SU_L(2) \times U(1) \times SU_f(3)$ Minkowski
space-time and what happens in it. The so-called "basic units of motion"
are made up from quarks (of six flavors, of three colors, and of
the two helicities) and leptons (of three generations and of the
two helicities). We use these basic units (of motion) in
writing down the lagrangian, etc. - the starting point of
our formalism(s).

If we look at the basic units of motion as compared to the original
particle, i.e. the electron, the starting basic units are all
"point-like" Dirac particles. Dirac invented the Dirac equation
for the electron eighty years ago and surprisingly enough these
"point-like" Dirac particles are the basic units of the
Standard Model. Thus, we call it "Dirac Similarity Principle"
- a salute to Dirac; a triumph to mathematics. Our world could
indeed be described by the proper mathematics. The proper
mathematical language may be the renormalizable quantum field
theory, as advocated in this paper.

For the lepton world or the quark world, the story is fixed if
the so-called "gauge-invariant derivative", i.e. $D_\mu$ in
the kinetic-energy term $-\bar \Psi \gamma_\mu D_\mu \Psi$, is
given for a given basic unit, one on one \cite{Books}.

For the lepton world, we introduce the family triplet,
$(\nu_\tau^R,\,\nu_\mu^R,\,,\nu_e^R)$ (column), under $SU_f(3)$.
Since the minimal Standard Model does not see the right-handed
neutrinos, it would be a natural way to make an extension of the
minimal Standard Model. Or, we have, for $(\nu_\tau^R,\,
\nu_\mu^R,\,\nu_e^R)$,
\begin{equation}
D_\mu = \partial_\mu - i \kappa {\bar\lambda^a\over 2} F_\mu^a.
\end{equation}
and, for the left-handed $SU_f(3)$-triplet and $SU_L(2)$-doublet
$((\nu_\tau^L,\,\tau^L),\, (\nu_\mu^L,\,\mu^L),\, (\nu_e^L,\,e^L))$
(all columns),
\begin{equation}
D_\mu = \partial_\mu - i \kappa {\bar\lambda^a\over 2} F_\mu^a - i g
{\vec \tau\over 2} \cdot \vec A_\mu + i {1\over 2} g' B_\mu.
\end{equation}
The right-handed charged leptons form the triplet $\Psi_R^C(3,1)$ under
$SU_f(3)$, since it were singlets their common factor $\bar\Psi_L(\bar 3,2)
\Psi_R(1,1)\Phi(3,2)$ for the mass terms would involve the cross terms such as
$\mu\to e$.

The neutrino mass term assumes a new form \cite{HwangYan}:
\begin{equation}
i {h\over 2} {\bar\Psi}_L(3,2) \times \Psi_R(3,1) \cdot
{\tilde \Phi}(3,2) + h.c.,
\end{equation}
where $\Psi(3,i)$ are the neutrino triplet just mentioned above (with
the first label for $SU_f(3)$ and the second for $SU_L(2)$). The
cross (curl) product is somewhat new \cite{Family}, referring to
the singlet combination of three triplets in $SU(3)$. The Higgs field
${\tilde \Phi}(3,2)$ is new in this effort, because it carries
some nontrivial $SU_L(2)$ charge.

Note that, for charged leptons, the Standard-Model choice is ${\bar \Psi}(\bar 3,2)
\Psi_R^C(3,1) \Phi(1,2) +c.c.$, which gives three leptons an equal mass. But, in
view of that if $(\phi_1,\phi_2)$ is an $SU(2)$ doublet then $(\phi_2^\dagger,
-\phi_1^\dagger)$ is another doublet, we could form ${\tilde\Phi}(3,2)$
from the doublet-triplet $\Phi(3,2)$. The notations in
$\Phi(1,2)$, $\Phi(3,2)$, and $\Phi(3,1)$ should be consistent
and thus the ${\tilde \Phi}(3,2)$, used in the above equation,
should have the tilde operation, for the consistency in notations.

So, we have \cite{Hwang417}
\begin{equation}
i {h^C\over 2} {\bar\Psi}_L(3,2) \times \Psi_R^C(3,1) \cdot
\Phi(3,2) + h.c.,
\end{equation}
which gives rise to the imaginary off-diagonal (hermitian) elements
in the $3\times 3$ mass matrix, so removing the equal masses of the
charged leptons. Note that the couplings $h$, $h^C$, and $\kappa$
all are dimensionless.

The expressions for neutrino oscillations and the off-diagonal mass term
are in $i\epsilon_{abc}$, or curl-dot, product - it is allowed for
$SU(3)$. Note that such coupling has nothing to do with the
kinetic-energy term of the particle, though the coupling $h$
(and $h^c$) might be related to the gauge coupling $\kappa$.

We now turn our attention to the quark world, which our special
gauge-group Minkowski space-time supports. Thus, we have, for
the up-type right-handed quarks $u_R$, $c_R$, and $t_R$,
\begin{equation}
D_\mu = \partial_\mu - i g_c {\lambda^a\over 2} G_\mu^a -
i {2\over 3} g'B_\mu,
\end{equation}
and, for the rotated down-type right-handed quarks $d'_R$, $s'_R$,
and $b'_R$,
\begin{equation}
D_\mu = \partial_\mu - i g_c {\lambda^a\over 2} G_\mu^a -
i (-{1\over 3}) g' B_\mu.
\end{equation}

On the other hand, we have, for the $SU_L(2)$ quark doublets,
\begin{equation}
D_\mu = \partial_\mu - i g_c {\lambda^a\over 2} G_\mu^a - i g
{\vec \tau\over 2}\cdot \vec A_\mu - i {1\over 6} g'B_\mu.
\end{equation}

There are the standard way to generate mass for the various quarks.
For these quarks, we use the "old-fashion" way as in the Standard
Model, since quarks do not couple to the family Higgs fields. We
have, for the generation of the various quark masses,
\begin{eqnarray}
& G_1 {\bar U}_L(1,2) u_R {\tilde \Phi}(1,2) + G'_1 {\bar U}_L(1,2) d'_R
\Phi(1,2) + h.c. +\nonumber\\
& G_2 {\bar C}_L(1,2) c_R {\tilde \Phi}(1,2) + G'_2 {\bar C}_L(1,2) s'_R
\Phi(1,2) + h.c. +\nonumber\\
& G_3 {\bar T}_L(1,2) t_R {\tilde \Phi}(1,2) + G'_3 {\bar T}_L(1,2) b'_R
\Phi(1,2) + h.c.,
\end{eqnarray}
with the tilde's defined as before.

Again, all the couplings in the quark world are
dimensionless in the 4-dimensional Minkowski space-time.
Surprisingly, the natural scale for the quark world is
of fermi scales, which is five orders smaller that the
natural scale of the lepton world, of atomic scales.

It might be essential to realize that the dimensionless
couplings $g_c$, $g$, $g'$ (hence $\alpha$), and $\kappa$
(in the strengths of the fundamental interactions)
and the dimensionless mass parameters $h$, $h^C$,
$G_{1,2,3}$, $G'_{1,2,3}$ (in describing the masses
of point-like particles) have a complete equal status
in the philosophy of concepts.

The overall background, i.e., the quantum 4-dimensional
Minkowski space-time with the force-fields gauge-group
structure $SU_c(3) \times SU_L(2) \times U(1) \times
SU_f(3)$ built-in from the outset, supports the
"dimensionless" lepton world and it also supports the
"dimensionless" quark world. It seems that there might
be many elegant stories associated with the Standard
Model (of all centuries) \cite{Hwang417}.

\medskip

\section{The Overall Background}

We should present our reasonings which lead to the formulation
of "The Origin of Mass" \cite{Origin}. It stresses that,
before the spontaneous symmetry breaking (SSB), the Standard Model
does not contain any parameter that is pertaining to "mass", but, after the SSB,
all particles in the Standard Model acquire the mass terms as it should - a way
to explain "the origin of mass". In this way, we sort of tie "the
origin of mass" to the effects of the SSB, or the generalized Higgs mechanism.

It is amusing to observe that it is so easy, by construction, to have
the complex scalar fields but, among the building blocks of matter, the
complex scalar fields are so rare. It seems to be much harder for two
complex scalar fields in co-existence, as though they would be mutually
"repulsive". Only if they belong to the "same" family, they might be
mutually attractive.

In the 4-dimensional Minkowski space-time, it is an amusing fact
for the complex scalar field that the dimensionless interaction
$\lambda (\phi^\dagger \phi)^2$ exists - we don't know how to
determine the dimensionless $\lambda$; this might have to
do with the 4-dimensional nature and maybe more. The
determination of $\lambda$, that should be done {\it a priori}
in the Standard Model, poses an important conceptional
question.

The reason that we try to write together a force-field Minkowski space-time
is that when put together the $SU_c(3) \times SU_L(2) \times U(1)
\times SU_f(3)$ Minkowski space-time (that is already specific
enough) the complex scalar field $\phi$ in this space should
have a specific $\lambda$ in the dimensionless interaction
$\lambda (\phi^\dagger \phi)^2$. If we agree that a specific
$\lambda$ is needed, then there is the universal $\lambda$ for
the various complex scalar fields allowed in this force-field
Minkowski space-time. The complex scalar field(s) should have
the existence {\it a priori}. [These arguments sound fairly
philosophical and logical, but they are needed for clarification.]

A complex scalar field in our space-time has the dimensionless coupling:
\begin{equation}
V(x) =  \lambda (\phi^\dagger(x) \phi(x))^2.
\end{equation}
The space-time integral of $L=T-V$ gives the action. In our 4-dimensional
Minkowski space-time, we find that $\lambda={1\over 8}$ numerically. This
number should come out topologically (after the normalizations of
the various fields in a given space \cite{Books}), although, at this
point, we don't know why this is the case.

If there are more than a complex scalar field, we should have
\begin{equation}
V(x) = \lambda \{(\phi_a^\dagger\phi_a)^2 + (\phi_b^\dagger
\phi_b)^2+ ....\}.
\end{equation}
There should be only one $\lambda$.

For the two related complex fields, we propose to write
\begin{equation}
V(x) = \lambda \{ (\phi_a^\dagger\phi_a + \phi_b^\dagger
\phi_b)^2 - 4 (\phi^\dagger_a \phi_b) \cdot (\phi^\dagger_b \phi_a)\},
\end{equation}
to signify the mutual attraction on top of the universal
repulsive interactions.

Now we return to the Standard Model of All Centuries \cite{Hwang417}.
We have the Standard-Model Higgs $\Phi(1,2)$, the purely family Higgs
$\Phi(3,1)$, and the mixed family Higgs $\Phi(3,2)$, with the first label
for $SU_f(3)$ and the second for $SU_L(2)$. We need another triplet $\Phi(3,1)$
since all eight family gauge bosons are massive \cite{Family}.

It is clear that $\Phi(1,2)$ would interact with $\Phi(3,2)$ while
$\Phi(3,1)$ would also interact with $\Phi(3,2)$. These interactions
should be attractive to explain why they are showing up
together.

We try to come back to the ground-zero point. Thus, we may
write down the general terms for
potentials among the three Higgs fields, subject to (1) that
they are renormalizable, and (2) that symmetries are only
broken spontaneously (the Higgs or induced Higgs mechanism).
We have \cite{Hwang417, Books}

\begin{eqnarray}
V = & V_{SM} +  V_1 + V_2 + V_3,\\
V_{SM} =& \mu^2 \Phi^\dagger(1,2) \Phi(1,2) + \lambda
(\Phi^\dagger(1,2) \Phi(1,2))^2\\
V_1 =& M^2 \Phi^\dagger(\bar 3,2) \Phi(3,2) + \lambda_1
(\Phi^\dagger(\bar 3,2) \Phi(3,2))^2\nonumber\\
  &+  \epsilon_1(\Phi^\dagger(\bar 3,2)\Phi(3,2))(\Phi^\dagger(1,2)\Phi(1,2))
   + \eta_1 (\Phi^\dagger(\bar 3,2)\Phi(1,2))(\Phi^\dagger(1,2)\Phi(3,2))
  \nonumber\\
  &+ \epsilon_2(\Phi^\dagger(\bar 3,2)\Phi(3,2))(\Phi^\dagger(\bar 3,1)\Phi(3,1))
   + \eta_2 (\Phi^\dagger(\bar 3,2)\Phi(3,1))(\Phi^\dagger(\bar 3,1)\Phi(3,2))
   \nonumber\\
  & + (\delta_1 i \Phi^\dagger(3,2)\times \Phi(3,2) \cdot \Phi^\dagger(3,1) +h.c.),\\
V_2 =& \mu_2^2 \Phi^\dagger(\bar 3,1)\Phi(3,1) + \lambda_2
(\Phi^\dagger(\bar 3,1) \Phi(3,1))^2 \nonumber\\
& + (\delta_2 i \Phi^\dagger(3,1)\cdot
\Phi(3,1) \times \Phi^\dagger(3,1) +h.c.)\nonumber\\
 &+ \lambda_2^\prime\Phi^\dagger(\bar 3,1) \Phi(3,1) \Phi^\dagger(1,2) \Phi(1,2),\\
V_3 =& (\delta_3 i \Phi^\dagger(3,2)\cdot \Phi(3,2)\times
    (\Phi^\dagger(1,2)\Phi(3,2)) +h.c.)\nonumber\\
    &+ (\delta_4 i (\Phi^\dagger(3,2)\Phi(1,2))\cdot \Phi^\dagger(3,1)
    \times \Phi(3,1)+h.c.)\nonumber\\
    & + \eta_3 (\Phi^\dagger(\bar 3,2)\Phi(1,2)\Phi(3,1)+c.c.).
\end{eqnarray}
In doing the renormalization analysis of the three Higgs fields,
we realize that even if we start from the well-motivated lagrangian
such as Eq. (11), it might spill over to the more generalized
lagrangian such as Eqs. (12)-(16).

We might pay special attention to the
so-called "U-gauge" (unitary gauge). In the U-gauge, every
particle is a real particle (not a ghost). We find it to
be useful in the analysis of the situation with the
spontaneous symmetry breaking (SSB). For the quantum
4-dimensional Minkowski space-time with the force-fields
gauge-group structure $SU_c(3) \times SU_L(2) \times U(1)
\times SU_f(3)$ built-in from the very beginning (i.e.,
the overall background), we have, in the U-gauge, $W^\pm$, $Z^0$,
and eight massive family gauge bosons, one Standard-Model Higgs and
four neutral family Higgs (three mixed plus one pure).

Thus, we choose to have, in the U-gauge, as in
\cite{Origin},
\begin{equation}
\Phi(1,2)= (0,{1\over \sqrt 2} (v+\eta)),\,\, \Phi^0(3,2) = {1\over \sqrt 2} (u_1+\eta'_1, u_2+
\eta'_2, u_3+\eta'_3 ),\,\, \Phi(3,1) = {1\over \sqrt 2}(w+\eta',0,0),
\end{equation}
all in columns. The five components of the complex triplet $\Phi(3,1)$ get
absorbed by the $SU_f(3)$ family gauge bosons and the neutral part of
$\Phi(3,2)$ has three real parts left - together making all eight family
gauge bosons massive.

Before the mixing, the masses of the various Higgs are given by, for
Eqs. (12)-(16),
\begin{eqnarray}
\eta:& (\mu^2/\lambda) + {1\over 4}(\epsilon_1+ \eta_1) u_i u_i + {\lambda'_2\over 4}
w^2, \nonumber\\
\eta':& (\mu_2^2/\lambda_2) + {\epsilon_2\over 4} u_i u_i +{\eta_2\over 4} u_1^2 +
{\lambda'_2\over 4} v^2,
\nonumber\\
\eta'_1:& M^2 + {1\over 4}(\epsilon_1+ \eta_1) v^2 +{1\over 4}(\epsilon_2 + \eta_2) w^2 +
(\lambda_1-term),\nonumber\\
\eta'_{2,3}:& M^2 + {1\over 4}(\epsilon_1+ \eta_1) v^2 + {\epsilon_2\over 4} w^2
+ (\lambda_1-term),\nonumber\\
\phi_1:& M^2 + {1\over 2}\epsilon_1 v^2 + {1\over 2}\epsilon_2 w^2 + {1\over 2}\eta_2 w^2 +
{\lambda_1\over 2} u_i u_i, \nonumber\\
\phi_{2,3}:& M^2 + {1\over 2}\epsilon_1 v^2 + {1\over 2} \epsilon_2 w^2
+ {\lambda_1 \over 2} u_i u_i.
\end{eqnarray}
The mixing term looks like, apart from some common factor:
\begin{equation}
2 (\epsilon_1+\eta_1)u_i\eta'_i v \eta + 2\epsilon_2 u_i\eta'_i w \eta' +
2 \eta_2 u_1\eta'_1 w\eta' + 2\lambda'_2 w\eta' v\eta.
\end{equation}
And we also neglect the mixing (and the mixing inside
$\eta'_{1,2,3}$). To understand "the origin of mass", we would
drop out all "mass" terms to begin with.

To understand the origin of mass \cite{Origin}, we find that the
ignition term would better be in the purely family sector, i.e.,
the $\mu_2^2$ term. When $\mu_2^2 = 0$, the $\Phi(3,2)$ is equally
partitioned between $\Phi(1,2)$ and $\Phi(3,1)$.

It is easy to see that only one SSB-driving term is enough for
all the three Higgs fields -- there may be several SSB's for
the neutral fields - in our case, it works for all of them. SSB for
one Higgs but is driven by other Higgs - a unique feature for the
complex scalar fields. Or, we have \cite{Origin}

\begin{eqnarray}
V_{Higgs} =& \mu^2_2 \Phi^\dagger(3,1) \Phi(3,1) + \lambda
(\Phi^\dagger(1,2) \Phi(1,2)+ cos\theta_P\Phi^\dagger(3,2)\Phi(3,2))^2\nonumber\\
    &  + \lambda(-4 cos\theta_P)
(\Phi^\dagger(\bar 3,2)\Phi(1,2))(\Phi^\dagger(1,2)\Phi(3,2))
  \nonumber\\
  &+\lambda
(\Phi^\dagger(3,1) \Phi(3,1)+ sin\theta_P \Phi^\dagger(3,2)\Phi(3,2))^2
\nonumber\\ & + \lambda(-4 sin\theta_P)
(\Phi^\dagger(\bar 3,2)\Phi(3,1))(\Phi^\dagger(3,1)\Phi(3,2)).
\end{eqnarray}
These are two prefect squares minus the other extremes, to guarantee
the positive definiteness, when the minus $\mu^2_2$ was left out.
($\theta_P$ may be referred to as "Pauchy's angle".) $\epsilon_1$,
$\eta_1$, and $\epsilon_2$, $\eta_2$ are expressed in $\lambda$, a
great simplification. Note that we only include the interference
terms between those involving the same group, $SU_f(3)$ or $SU_L(2)$;
thus $\lambda'_2 =0$.

From the expressions of $u_iu_i$ and $v^2$, we obtain
\begin{equation}
v^2 (3 cos^2\theta_P-1) = sin\theta_P cos\theta_P w^2.
\end{equation}
And the SSB-driven $\eta'$ yields
\begin{equation}
w^2 (1-2 sin^2\theta_P) = - {\mu_2^2\over \lambda} +
(sin2\theta_P - tan\theta_P) v^2.
\end{equation}
These two equations show that it is necessary to have the driving
term, since $\mu^2_2=0$ implies that everything is zero. Also,
$\theta=45^\circ$ is the (lower) limit.

The mass squared of the SM Higgs $\eta$ is $2\lambda cos\theta_P u_i u_i$,
as known to be $(125\,\, GeV)^2$. The famous $v^2$ is the number
divided by $2\lambda$, or $(125\,\,GeV)^2/(2\lambda)$. Using PDG's for
$e$, $sin^2\theta_W$, and the $W$-mass \cite{PDG}, we find
$v^2=255\,\, GeV$. So, $\lambda={1\over 8}$, a simple model indeed.

The ratio of the VEV to its Higgs mass is determined
by $2\lambda$, whether the channel is not ignited or not. We might
choose the channel of $\eta'$ (the purely family Higgs) or that of
$\eta$ (the SM Higgs) as the ignition channel, but three Higgs channels
have different labels. The three Lorentz-invariant scalar fields have
different internal structures - an amusing question for further
investigation.

The mass squared of $\eta'$ is $-2(\mu_2^2-sin\theta_P u_1^2 +
sin\theta_P (u_2^2+u_3^2))$. The  other condensates are $u_1^2= cos\theta_P v^2
+ sin\theta_P w^2$ and $u_{2,3}^2 = cos\theta_P v^2 - sin\theta_P w^2$ while
the mass squared of $\eta'_1$ is $2\lambda u_1^2$, those of $\eta'_{2,3}$ be
$2\lambda u_{2,3}^2$. The mixings among $\eta'_i$ themselves are neglected
in this paper.

There is no SSB for the charged Higgs $\Phi^+(3,2)$. The mass
squared of $\phi_1$ is $\lambda(cos\theta_P v^2 - sin\theta_P w^2) + {\lambda\over 2}
u_i u_i$ while $\phi_{2,3}$ be $\lambda(cos\theta_P v^2 + sin\theta_P w^2)
+ {\lambda\over 2} u_i u_i$. (Note that a factor of ${1\over 2}$ appears
in the kinetic and mass terms when we simplify from the complex case to
that of the real field; see Ch. 13 of the Wu-Hwang book \cite{Books}.)

A further look of these equations tells that $3cos^2\theta_P - 1 > 0$ and
$2sin^2\theta_P -1 > 0$. A narrow range of $\theta_P$ is allowed (greater
than $45^\circ$ while less than $57.4^\circ$, which is determined by
the group structure). For illustration, let us choose
$cos \theta_P = 0.6$ and work out the numbers as follows:
(Note that $\lambda={1\over 8}$ is used.)
\begin{eqnarray}
& 6 w^2 = v^2, \quad -\mu^2_2/\lambda = 0.32 v^2;\nonumber\\
\eta: & 2\lambda cos\theta_0 u_i u_i =(125\, GeV)^2, \quad v^2 = (250\,GeV)^2;
\nonumber\\
\eta': & mass^2 = (51.03\,GeV)^2, \quad w^2=v^2/6; \nonumber\\
\eta'_1: & mass^2= (107\,GeV)^2, \quad u_1^2=0.7333 v^2; \nonumber\\
\eta'_{2,3}: & mass^2 = (85.4\,GeV)^2, \quad u_{2,3} = 0.4667 v^2; \nonumber\\
\phi_1:& mass = 100.8\, GeV; \qquad \phi_{2,3}: mass = 110.6\,GeV.
\end{eqnarray}
All numbers appear to be reasonable. In the above, $cos\theta_P$ is the
only free parameter until one of the family Higgs particles $\eta'_{1,2,3}$
and $\eta'$ is found experimentally. Since the new objects need to be
accessed in the lepton world, it would be a challenge for our experimental
colleagues.

{\it As a footnote, our Standard Model predicts that the mass of the SM
Higgs $\eta$ is a half of the vacuum expectation value $v$ - a prediction
in the origin of mass \cite{Origin}.}

As for the range of validity, ${1\over 3} \le cos^2\theta_P \le {1\over 2}$.
The first limit refers to $w^2=0$ while the second for $\mu_2^2 = 0$.

We may fix up the various couplings, using our common senses. The
cross-dot products would be similar to $\kappa$, the basic coupling of
the family gauge bosons. The electroweak coupling $g$ is
$0.6300$ while the strong QCD coupling $g_s=3.545$; my first guess
for $\kappa$ would be about $0.1$. The masses of the family gauge
bosons would be estimated by using ${1\over 2}\kappa \cdot w$, so
slightly less than $10\,GeV$. (In the numerical example with $cos
\theta_P=0.6$, we have $6 w^2= v^2$ or $w=102\,\,GeV$. This gives
$m=5\,\,GeV$ as the estimate.) So, the range of the family forces,
existing in the lepton world, would be $0.02\,\, fermi$.

In \cite{Origin}, the term that ignites the SSB is chosen to be with
$\eta'$, the purely family Higgs. This in turn ignites EW SSB
and others. It explains the origin of all the masses, in terms
of the spontaneous symmetry breaking (SSB). SSB in $\Phi(3,2)$
is driven by $\Phi(3,1)$, while SSB in $\Phi(1,2)$ from the
driven SSB by $\Phi(3,2)$, as well. The different, but related,
scalar fields can accomplish so much, to our surprise.

We note that, at the Lagrangian level, the $SU_c(3) \times SU_L(2)
\times U(1) \times SU_f(3)$ gauge symmetry is protected but the symmetry
is violated via spontaneous symmetry breaking (via the Higgs mechanisms).

We iterate that the mathematics of the three neutral Higgs, $\Phi(1,2)$
(Standard-Model Higgs), $\Phi(3,1)$ (purely family Higgs), and
$\Phi^0(3,2)$ (mixed family Higgs), subject to the renormalizabilty
(up to the fourth power), turns to be rather rich. In our earlier work
regarding the "colored Higgs mechanism" \cite{HwangWYP}, we show how
the eight gauge bosons in the $SU(3)$ gauge theory become massive using two
complex scalar triplet fields (with the resultant four real Higgs fields),
with a lot of choices. We suspect that, even within QCD, there might be
some elegant choice of "colored" Higgs, or there must be a good reason
for massless gluons.

\medskip

\section{The Standard Model as a complete theory}

We declare that we are living in the quantum 4-dimensional
Minkowski space-time with the force-fields gauge-group
structure $SU_c(3) \times SU_L(2) \times U(1) \times
SU_f(3)$ built-in from the very beginning. This "overall
background" can see the lepton world, of atomic sizes,
that has the $SU_L(2) \times U(1) \times SU_f(3)$ symmetry,
or the other (123) symmetry. It also can see the quark world
of much smaller nuclear sizes, which possesses the
$SU_c(3) \times SU_L(2) \times U(1)$ symmetry, or the
standard (123) symmetry.

We make this declaration so that we spell out
an end to Newton's classic era of four centuries. It is
fundamental to recognize that we are living in a "new"
world, that is different from the classic Newton's
world.

The lepton world is dimensionless in the 4-dimensional
Minkowski space-time. The quark world is also dimensionless
in the 4-dimensional Minkowski space-time. Except the
SSB "ignition" term, the overall background is also
dimensionless in the 4-dimensional Minkowski space-time.
"Dimensionless in the 4-dimensional Minkowski space-time"
means that it is determined {\it globally} by the
quantum 4-dimensional Minkowski space-time.

To be precise, the lepton world behaves well in the
quantum 4-dimensional Minkowski space-time built-in
with the force-fields gauge-group structure $SU_L(2)
\times U(1) \times SU_f(3)$. Meanwhile, the quark
world behaves well in the quantum 4-dimensional
Minkowski space-time built-in with the force-fields
gauge-group structure $SU_c(3) \times SU_L(2) \times
U(1)$. All the couplings are dimensionless and, thus,
they are determined {\it globally} by the quantum
4-dimensional Minkowski space-time.

Can we implement some methodology of ultraviolet
divergences such that everything is determined {\it globally}
by the quantum 4-dimensional Minkowski space-time with some
force-fields gauge-group structure built-in from the very
beginning? As we shall see, it is rather difficult to obtain
clear-cut answers to these questions.

The lepton world respects the $SU_L(2) \times U(1)
\times SU_f(3)$ symmetry such that, in principle,
the divergences in the lepton world should "operate"
among themselves. Similar cases happen for the
quark world. Likewise, those in the overall
background should also "operate" among themselves.
The mix-ups among the different sectors occur so
easily if we go to higher "orders".

Let's try to think following a specific
example. We consider the the self-energy
of the Standard-Model (SM) Higgs $\Phi(1,2)$,
depicted by Figs. 1. It is obvious that
there are mix-ups among the different sectors.

Specifically, for the sake of simplicity, we are
focusing our attention only on the ultraviolet
divergences of highest order in one loop.

Hence we could only begin the analysis of "one" such question
while leaving the heavy burdens to the others. In the
textbook, e.g., Ch. 10 of the Wu-Hwang book \cite{Books}
on the ultraviolet divergences in QED, the divergences
are there, but QED is only part of the theory, including
the leading-order calculation related to $g-2$. We propose
that many issues could be examined in a theory in the
quantum 4-dimensional Minkowski space-time with the
force-fields gauge-group structure $SU_c(3) \times
SU_L(2) \times U(1) \times SU_f(3)$ built-in from the
very beginning.

Although one electron self-energy diagram shown in
Ch. 10 of the Wu-Hwang book \cite{Books} is infinite, this is
true in QED but QED is {\it only} part of the Standard Model.
Maybe all of the self-energy diagrams of similar type could
add up to a finite number. The fact that the QED-part is
infinite is the symptom of that QED is an incomplete theory.
(One can argue this way.) In a compete theory, we
should have everything, and nothing more - if there is still
some infinities, then there is something more; this is a "new"
way of searching for new things, i.e., the hard way, to find
something new.

In fact, we would like to show that the U-gauge and the dimensional
regularization offer us the {\it machinery} to handle these
infinities. Staying in the U-gauge means that we are treating
"physical particles". Dimensional regularization means that the
results in all dimensions are included, including finite results
in fractional dimensions.

But, unfortunately, the dimensional regularization
{\it does not care about the causality $i\epsilon$
requirements} so that the results are only telling
us something, and they may not be the real final
numbers. In fact, the answers which we obtain in
the dimensional regularization scheme may not be
{\it true} answers (under the casual $i\epsilon$
prescription), since in the dimensional
regularization they have their own definition
of the fractional-dimensional integrals.

On the negative side, in the dimensional
regularization scheme, one tries to define the
results for the fractional dimensions and, by
continuation, obtain the results for the integral
dimensions. Unlike the Pauli-Villars regularization
(e.g., Ch. 10 of \cite{Books}), one tries to
manipulate the causality $i\epsilon$ requirement
in obtaining the final results.

We try to follow the details of \cite{Origin} in
discussing the ultraviolet divergences of the
quadratic order of the self-energy of the SM
Higgs, but adding some new points.

\begin{figure}[h]
\centering
\includegraphics[width=4in]{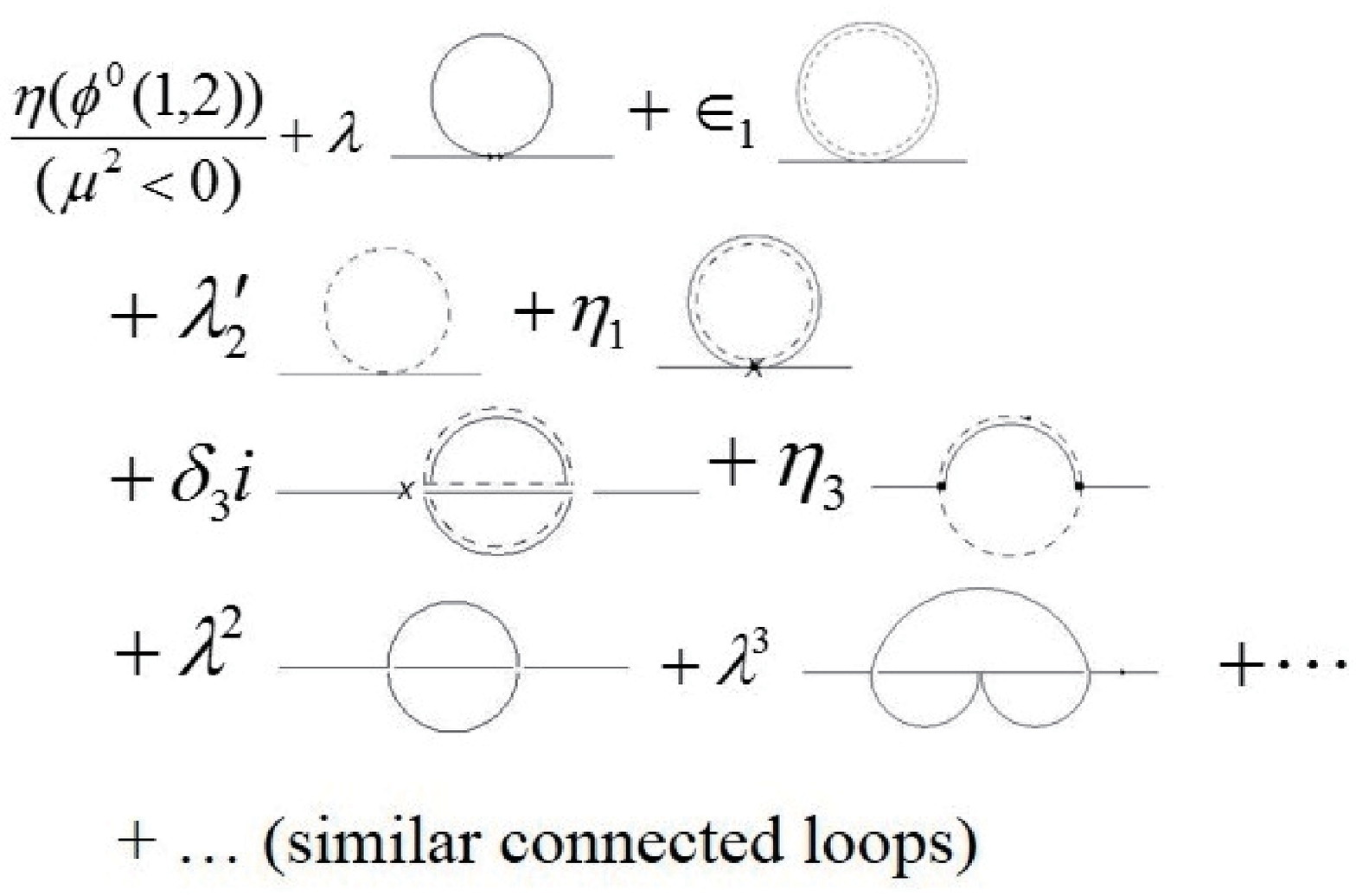}
\caption{The within-Higgs diagrams for the Standard-Model Higgs $\Phi(1,2)$.}
\end{figure}

In Fig. 1, the wave-function renormalization of the Standard-Model Higgs
$\Phi(1,2)$ is shown, for simplicity, in the U-gauge in the absence of Dirac
fermions. The lowest-order loop diagrams, from the above interaction
lagrangian, are shown from 1(b) [in $\lambda$] to 1(g) [in $\eta_3$],
where the first five are of quadratic divergence while the last one of
logarithmic divergence. The higher-order connected loop diagrams, many
of them and also of quadratic divergence, are also troublesome and should
be dealt with at the same time. We will discuss the cases of the worst
divergent, i.e., the quadratic divergent, in what follows.

Using dimensional regularization (i.e. the appendix of Ch. 10, the
Wu-Hwang book, Ref. \cite{Books}), we could write down the
one-loop results.

We try to use one explicit example to illustrate our point related to
the infinities - the quadratic divergences of the wave function of the
SM Higgs $\eta$. Here, in Fig. 1, we try to show only the Higgs sectors
themselves; in a complete Standard Model, the (Dirac) fermion
loop diagrams, and those with gauge bosons, also present divergences
of quadratic order and should be dealt with simultaneously.

As said earlier, we know that the formulae in dimensional
regularization give us some things, {\it apart from the
$-i\epsilon$ prescription}, and that it "works" in the
U-gauge. For example, the $Z^0-boson$ loop for Fig. 1
would give us the vanishing result - so, it does not
bother us.

In details, the coupling of the SM Higgs is (\cite{Books}, e.g.,
Wu/Hwang, Ch. 13, Ref. \cite{Books})
\begin{equation}
-{1\over 8}(v^2+2v\eta +\eta^2) \{2g^2 W_\mu^+W_\mu^-
+ [g^2+(g')^2] Z_\mu^0 Z_\mu^0\},
\end{equation}
which gives rise to, to the first order, the one-loop $W^\pm$
or $Z^0$ diagram. To evaluate them, we use the propagator
in the U-gauge (see the appendix of Ch. 13, Ref. \cite{Books})
and the formulae in the dimensional regularization (see the
appendix of Ch. 10, Ref. \cite{Books}). They cancel between
two terms for each diagram.

We proceed to examine those diagrams in Fig. 1 which are
"simple" quadratically divergent - those at the one-loop
order. These are among the various Higgs.

The one-loop diagrams involving the quark (or charged lepton),
when simplified, are sums of quadratic and logarithmic
divergences.

Using dimensional regularization (i.e. the appendix of Ch. 10, the
Wu-Hwang book, \cite{Books}), we obtain the one-loop and quadratic-divergence
results as follows. In the dimensional regularization, the factor
$\Gamma(1-{n\over 2})$ stands for where the quadratic divergence
appears. Maybe the fractional dimensions, which are represented as
finite numbers, could get some meaning, but we have to remember
that, as a drawback, we bypass the $-i\epsilon$ in the propagators.

\begin{eqnarray}
& -4\cdot {n\over 2}\cdot (S_q + S_{c.l.})\Gamma(1-{n\over 2}) \nonumber\\
&+ \{ 3\lambda m^2(\eta) + {\epsilon_1\over 2} \sum_i m^2(\eta'_i)
+ \epsilon_1 \sum_i m^2(\phi_i)\nonumber\\
& +{\lambda_2^\prime\over 2} m^2(\eta') +
{\eta_1\over 2} \sum_i m^2(\eta'_i)\}\Gamma(1-{n\over 2});\\
& S_q = \sum_{quarks} 3\cdot G_i^2\cdot (m_i^2 -{1\over 6} m^2(\eta)),\nonumber \\
& S_{c.l.} = 3\cdot {\bar G}_l^2 \cdot ({\bar m}_l^2 -{1\over 6} m^2(\eta)).
\end{eqnarray}
Or, using the Standard Model of the paper, we have
\begin{eqnarray}
& -4\cdot {n\over 2}\cdot (S_q+S_{c.l.})\Gamma(1-{n\over 2})\nonumber\\
& + \{\lambda (3m^2 (\eta) - cos\theta_P\sum_i m^2(\eta'_i) + 2 cos\theta_P
\sum_i m^2(\phi_i)) \} \Gamma(1-{n\over 2}).
\end{eqnarray}
Here we don't equal the sum to anything since, as
we said before, the dimensional regularization
scheme don't take care of the causality $i\epsilon$
prescription.

Maybe we could proceed in the following way. After
removing the divergent $\Gamma(1-{n\over 2})$ parts,
we might propose to use
\begin{eqnarray}
& -4\cdot {n\over 2}\cdot (S_q+S_{c.l.})\nonumber\\
& + \{\lambda (3m^2 (\eta) - cos\theta_P\sum_i m^2(\eta'_i) + 2 cos\theta_P
\sum_i m^2(\phi_i)) \} \approx 0;
\end{eqnarray}
or something finite.

Here we would be happy to use an equal sign instead
of the approximate sign, if the causality $-i\epsilon$
prescription could be used instead.

There are a few general characteristics: (1) In the contributions
from quarks and from charged leptons, the mass of the SM Higgs enters
(as external momentum squared). This makes all contributions
equivalent in some sense. (2) We assume that the quarks enter in the
theory in the SM way - if we examine the theory closely, there might
still be some room for colored Higgs mechanism \cite{HwangWYP}.
In other words, we are not sure that the "identity" in this
SM Higgs would hold out; instead, the identity in the case of
pure Higgs $\Phi(3,1)$, or that for $\Phi(3,2)$, has better
reasons to hold out. Or, in light of the fact that it is
difficult to rule out the colored Higgs mechanism in the quark
world.

In deriving the above equations, the coefficients of
$\Gamma(1-{n\over 2})$ are the coefficients of
quadratic divergences while those of $\Gamma(2-{n\over 2})$
are the coefficients of logarithmic divergences - for the
latter, divergence is less severe and the contributions could
be everywhere; the treatment is far more complicated.

As mentioned in \cite{Origin}, the diagrams which are of multiple
quadratic divergence are troublesome since the series could be blown-up,
of $2n$-th divergence with $n \to \infty$. Mathematically, we should
avoid such terms by all means. This is the requirement beyond the
naive renormalizability.

We are hinting that we have to study the mathematics of divergences;
they are there, because of the uncountably infinite degrees of
freedom and other reasons, and there are regularities to be
discovered \cite{fine-tune}. So, is the Standard Model a complete
theory?

The result for $\Phi(1,2)$ in the one-loop result certainly gives
rise to an "approximate" formula for the masses and the couplings
- they are "approximate" because the cancelation could be modified
by terms in the higher-loop orders. But it should be approximate -
that is why it is worthwhile looking for them.

According to dimensional-regularization results, the three-loop
diagram gives rise to (quadratic divergence) $\times$ (logarithmic
divergence), and so on. They have to organized differently. One simple
way out is that they cancel completely in their own group, such as
all the four-loop diagrams. In any event, these divergences could
be systematically analyzed to display what is going on.

Maybe we have said too much about the dimensional-regularization
results, we in fact are suspicious that these results may not
make any sense since they ignore the causality $-i\epsilon$
requirement but the causality $-i\epsilon$ requirement is a
must. Let's look at the fermion-loop diagram for the
Higgs $\Phi(1,2)$, which is given by the previous
$S_{c.l.}$ term in the dimensional regularization scheme.

\bigskip

\begin{eqnarray}
& I(k^2) \qquad\qquad\qquad\qquad\qquad\qquad\qquad \nonumber\\
= & \int {d^4 p\over (2\pi)^4}
Tr\{ {m + i\gamma\cdot p\over m^2 +p^2 - i\epsilon}
{m - i\gamma\cdot (p+k) \over m^2 + (p+k)^2 -i \epsilon}
(1 - \gamma_5)\} \nonumber\\
= & 4 \int {m^2 + p\cdot (p+k) \over (m^2 +p^2 -i\epsilon)
(m^2+ (p+k)^2 -i\epsilon)}\nonumber\\
\equiv & (I(k^2) - I(0)) + I(0). 
\end{eqnarray}
Or, it can be represented as a Taylor's series.

In $I(0)$, it becomes a simple pole in the $k_0$ 
integration. In the next Taylor's expansion, integration 
by parts could help us to convert the double pole to 
a simple pole. So, the math tricks could make everything
transparent to us.

But, in light to the infinities, it is not obvious on 
how to proceed. What we are doing is a little similar to 
the Pauli-Villars regularization scheme, e.g., Ch. 10 of 
the Wu-Hwang book \cite{Books}, though we didn't
work out infinities. The dimensional regularization
scheme \cite{Books} ignores the causality $i\epsilon$
requirement; in our opinions, we don't know what we are 
doing. 

If we believe that all point-like particles are there and
that the Standard Model describes them so well, then some
sort of cancelation conjecture or hypothesis might hold,
first for the simple quadratic divergences of a certain
Higgs, and then other divergences. Of course some
theorem(s) of "global" type may be needed. The "new"
marriage of mathematics and physics is needed for the
ultimate breakthrough. Infinities in our language should 
not discourage us since the degrees of freedom in quantum 
fields are more than enumerable infinite.

\medskip

\section{Other Important Observations}

It is very strange that our overall background
is the quantum 4-dimensional Minkowski space-time
with the force-fields gauge-group structure
$SU_c(3) \times SU_L(2) \times U(1) \times SU_f(3)$
built-in from the very beginning. The uniform
$3^\circ\,K$ cosmic microwave background (CMB)
serves as the best evidence of this overall
background.

Moreover, this overall background sees the
lepton world, or the various atomic worlds.
The couplings are all dimensionless in the
4-dimensional Minkowski space-time.

And, moreover, this overall background sees
the quark world, or the various nuclear worlds.
Again, the couplings are all dimensionless in
the 4-dimensional Minkowski space-time.

These "dimensionless-ness" should enter our
discussions of ultraviolet divergences. But we
don't know how - maybe we anticipate to have
a good starting point to construct the Standard
Model of All Centuries.

Why do we need the "ignition" channel? Why is the
ignition channel not the SM Higgs $\eta$? We set out
to use the SM Higgs as the "ignition" channel, but
soon realized that the "ignition" channel with the
purely family $\eta'$ works out "perfectly". In fact,
the first version of \cite{Origin} was based on that
the ignition channel was the SM Higgs $\eta$ - the standard
wisdom. As for why we need the "ignition" channel, we still
need a good answer. (We believe that the God would not choose
more than one ignition point, in this game.)

Note that $\theta_P=45^\circ$ corresponds to the situation
that $\Phi(3,2)$ be equally divided by $\Phi(1,2)$
(SM Higgs) and $\Phi(3,1)$ (purely family). The situation
corresponds to that it is not yet "ignited" ($\mu_2^2$ =0).
How do these translate into the temperature situations
(via the Big Bang or via big inflation)?
The elusive purely family Higgs $\eta'$ as the "ignition"
channel gives us a few interesting questions.

\medskip

\section{Concluding Remarks}

To close this paper, we append a few remarks just
to remind ourselves the leading physical issues that we
may pursue after.

To verify this Standard Model is the experimental search for the
family Higgs $\eta'_1$, or $\eta'_{2,3}$, or charged family
Higgs $\phi_1^+$ and $\phi^+_{2,3}$, or pure family Higgs
$\eta'$, in a proposed $120\,GeV$ $\mu^+e^-$ collider
\cite{collider}.

The major implication of the family gauge theory is in fact a multi-GeV or
sub-sub-fermi gauge theory (a new force field of a few $10^{-15}\, cm$
in the range), assuming that the ordering in the coupling
constants, $g_W/ g_c \sim  \kappa/g_W$, is reasonable. Note that
the lepton world are shielded from this $SU_f(3)$ theory against the QED
Landau's ghost, and similarly the quark world from strong-interaction
$SU_c(3)$. The $g-2$ anomaly certainly deserves another serious look
in this context \cite{Kinoshita}.

In this Standard Model, the masses of quarks are
diagonal, or the singlets in the $SU_f(3)$ space; those of the
three charged leptons are $m_0 + a \lambda_2 + b \lambda_5 +
c \lambda_7$ (before diagonalization) and the masses of neutrinos are
purely off-diagonal, i.e. $a' \lambda_2 + b' \lambda_5 + c' \lambda_7$.
This result is very interesting and very intriguing.

This result follows from the above curl-dot product, or, the $\epsilon^{abc}
\bar \Psi_{L,a} \Psi_{R,b} \Phi_c$ product, i.e. the
$SU_f(3)$ operation, in writing the coupling(s) to the right-handed
lepton triplets.

In addition, neutrinos oscillate among themselves, giving rise to a
lepton-flavor-violating interaction (LFV). There are other
oscillation stories, such as the oscillation
in the $K^0-{\bar K}^0$ system, but there is a fundamental "intrinsic"
difference here - the $K^0-{\bar K}^0$ system is composite while neutrinos
are "point-like" Dirac particles. We have standard Feymann diagrams for
the kaon oscillations but similar diagrams do not exist for point-like neutrino
oscillations - our Standard Model solves the problem, maybe in a unique way.

Thinking it through, it is true that neutrino masses and neutrino
oscillations may be regarded as one of the most important experimental
facts over the last thirty years \cite{PDG}.

In fact, certain LFV processes such as $\mu \to e + \gamma$ \cite{PDG},
$\mu + A \to A^* + e$, etc., are closely related to the most cited
picture of neutrino oscillations \cite{PDG}. In our previous
publications \cite{Hwang10}, it was pointed out that the cross-generation
or off-diagonal neutrino-Higgs interaction may serve as the detailed mechanism
of neutrino oscillations, with some vacuum expectation values of the family
Higgs $\Phi^0(3,2)$. So, even though we haven't seen, directly,
the family gauge bosons and family Higgs particles, we have already seen the
manifestations of their vacuum expectation values.

Moreover, in this Standard Model, neutrinos and antineutrinos
are the {\it only long-lived} dark-matter particles \cite{HwangP1}.
Cosmic background $\nu$'s from the early Universe, owing to the
nonzero neutrino masses, would cluster in lumps, in neutrino
halos, five times in weight the visual objects, such as stars,
planets, etc. Since the neutrino halo is incompressible
(as a Fermi gas), this would prevent the final collapse of the
visual ordinary-matter object into a black hole \cite{neutrino}.

Thus, the Standard Model, i.e., the quantum 4-dimensional
Minkowski space-time with the force-fields gauge-group
structure $SU_c(3) \times SU_L(2) \times U(1)
\times SU_f(3)$ built-in from the very beginning, we
understand our Universe; that is, all the dark-matter
particles and all the ordinary-matter particles are
accounted for. Formation of the black holes for visual
ordinary-matter objects are stopped by the neutrino
halos (of the five times in weight).  Our Standard Model
gives us the perfect Universe which we are living.


\medskip

\begin{thebibliography}{99}

\bibitem{Hwang417} W-Y. Pauchy Hwang, arXiv:1304.4705v2 [hep-ph]
25 Aug 2015; "The Standard Model".

\bibitem{Origin} W-Y. Pauchy Hwang, The Universe, {\bf 2-2}, 47 (2014);
"The Origin of Mass".

\bibitem{HwangYan} W-Y. Pauchy Hwang and Tung-Mow Yan, The Universe,
Vol. 1, No. 1, 5 (2013); arXiv:1212.4944v1 [hep-ph] 20 Dec 2012.

\bibitem{Family} W-Y. Pauchy Hwang, Nucl. Phys. {\bf A844}, 40c (2010);
W-Y. Pauchy Hwang, International J. Mod. Phys.
{\bf A24}, 3366 (2009); the idea first appeared in
hep-ph, arXiv: 0808.2091; talk presented at 2008 CosPA Symposium
(Pohang, Korea, October 2008), Intern. J. Mod. Phys. Conf. Series {\bf 1}, 5
(2011); plenary talk at the 3rd International Meeting on Frontiers of Physics,
12-16 January 2009, Kuala Lumpur, Malaysia, published in American Institute of
Physics 978-0-7354-0687-2/09, pp. 25-30 (2009).

\bibitem{Books} Ta-You Wu and W-Y. Pauchy Hwang, "Relatistic Quantum
Mechanics and Quantum Fields" (World Scientific 1991); Francis Halzen
and Alan D. Martin, "Quarks and Leptons" (John Wiley and Sons, Inc.
1984); E.D. Commins and P.H. Bucksbaum, "Weak Interactions of Leptons
and Quarks" (Cambridge University Press 1983). We use the first book
for the notations and the metric(s).

\bibitem{HwangWYP} W-Y. P. Hwang, Phys. Rev. {\bf D32}, 824 (1985).

\bibitem{fine-tune} W-Y. Pauchy Hwang, The Universe, {\bf 2-1}, 41
(2014).

\bibitem{collider} W-Y. Pauchy Hwang, arXiv:1409.6296v1 [hep-ph]
22 Sep 2014; "The Family Collider".

\bibitem{Kinoshita} T. Kinoshita and W.B. Lindquist, Phys. Rev. Lett.
{\bf 47}, 1573 (1981).

\bibitem{PDG} Particle Data Group, "Review of Particle Physics",
J. Phys. G: Nucl. Part. Phys. {\bf 37}, 1 (2010); and its biennual
publications.

\bibitem{Hwang10} W-Y. Pauchy Hwang, Hyperfine Interactions {\bf 215}, 105 (2013).

\bibitem{HwangP1} W-Y. Pauchy Hwang, arXiv:1012.1082v4 [hep-ph] 13 Jan
2016; "Cosmology: Neutrinos as the Only Final Dark Matter".

\bibitem{neutrino} W-Y. Pauchy Hwang, The Universe, {\bf 3-4}, 7 (2015).

\end{thebibliography}
\end{document}